\documentclass[aps,prl,reprint,groupedaddress,floatfix,superscriptaddress,amsmath,showpacs,longbibliography]{revtex4-1}

\usepackage{graphicx}% Include figure files
\usepackage{dcolumn}% Align table columns on decimal point
\usepackage{bm}% bold math
\usepackage[colorlinks=true,linkcolor = blue,citecolor=blue,urlcolor=blue,anchorcolor = blue]{hyperref}
\usepackage{wasysym}
\usepackage{stmaryrd}
\usepackage{verbatim}
\usepackage{subfigure}
\usepackage{amsmath}
\usepackage{times}
\usepackage[version=4]{mhchem}
\usepackage{braket}
\usepackage{siunitx}
\usepackage{siunitx}
\usepackage{amssymb}
\newcommand{\RNum}[1]{\uppercase\expandafter{\romannumeral #1\relax}}

\begin{document}
	
\title{Effects of the Crystalline Electric Field in the \ce{KErTe2} Quantum Spin Liquid Candidate}

\author{Weiwei\,Liu}
\thanks{These authors contributed to the work equally.}
\affiliation{Department of Physics, Renmin University of China, Beijing 100872, China}
\affiliation{Beijing National Laboratory for Condensed Matter Physics, Institute of Physics, Chinese Academy of Sciences, Beijing 100190, China}

\author{Zheng\,Zhang}
\thanks{These authors contributed to the work equally.}
\email[e-mail:]{ruc\_zhangzheng@ruc.edu.cn}
\affiliation{Department of Physics, Renmin University of China, Beijing 100872, China}
\affiliation{Beijing National Laboratory for Condensed Matter Physics, Institute of Physics, Chinese Academy of Sciences, Beijing 100190, China}

\author{Dayu\,Yan}
\affiliation{Beijing National Laboratory for Condensed Matter Physics, Institute of Physics, Chinese Academy of Sciences, Beijing 100190, China}

\author{Jianshu\,Li}
\affiliation{Beijing National Laboratory for Condensed Matter Physics, Institute of Physics, Chinese Academy of Sciences, Beijing 100190, China}

\author{Zhitao\,Zhang}
\affiliation{Anhui Province Key Laboratory of Condensed Matter Physics at Extreme Conditions, High Magnetic Field Laboratory, Chinese Academy of Sciences, Hefei 230031, China}

\author{Jianting\,Ji}
\affiliation{Beijing National Laboratory for Condensed Matter Physics, Institute of Physics, Chinese Academy of Sciences, Beijing 100190, China}

\author{Feng\,Jin}
\affiliation{Beijing National Laboratory for Condensed Matter Physics, Institute of Physics, Chinese Academy of Sciences, Beijing 100190, China}

\author{Youguo\,Shi}
\affiliation{Beijing National Laboratory for Condensed Matter Physics, Institute of Physics, Chinese Academy of Sciences, Beijing 100190, China}

\author{Qingming\,Zhang}
\affiliation{School of Physical Science and Technology, Lanzhou University, Lanzhou 730000, China}
\affiliation{Beijing National Laboratory for Condensed Matter Physics, Institute of Physics, Chinese Academy of Sciences, Beijing 100190, China}

\begin{abstract}
Since discovering the \ce{ARECh2} (A=alkali or monovalent ions, RE=rare earth, Ch= chalcogen) triangular lattice quantum spin liquid (QSL) family, its oxide, sulfide, and selenide members have been continuously reported and extensively studied.
\ce{KErTe2} is the first synthesized telluride member and its spin triangular lattice remains unchanged. 
It was, however, expected that large tellurium ions could introduce more remarkable magnetic features and electronic structures in
this family of materials. 
In this paper, we performed thermodynamic and electron spin resonance (ESR) measurements to study low-energy magnetic excitations,
which were significantly affected by crystalline electric field (CEF) excitations due to relatively small gaps between the CEF ground state and the excited states.
Based on the CEF and mean-field (MF) theories, we analyzed systematically and consistently the ESR experiments and thermodynamic measurements including susceptibility, magnetization, and heat capacity.
The CEF parameters were successfully extracted by fitting high-temperature ($\textgreater$ 20 K) susceptibilities in the ab-plane and along the c-axis, allowing to determine the Lande factors ($g_{ab,calc}$ = 5.98(7) and $g_{c,calc}$ = 2.73(3)). These values were consistent with the values of Lande factors determined by ESR experiments ($g_{ab,exp}$ = 5.69 and $g_{c,exp}$ = 2.75).
By applying the CEF and MF theories to the susceptibility and magnetization results, we estimated the anisotropic spin-exchange energies and found that the CEF excitations in \ce{KErTe2} played a decisive role in the magnetism above 3 K, while the low-temperature magnetism below 10 K was gradually correlated with the anisotropic spin-exchange interactions. 
The CEF excitations were demonstrated in the low-temperature heat capacity, where both the positions of two broad peaks and their magnetic field dependence well corroborated our calculations. 
The present study provides a basis to explore the enriched magnetic and electronic properties of the QSL family.
\end{abstract}

\maketitle

%----------------------------
\begin{figure*}[t]
	\includegraphics[scale=1]{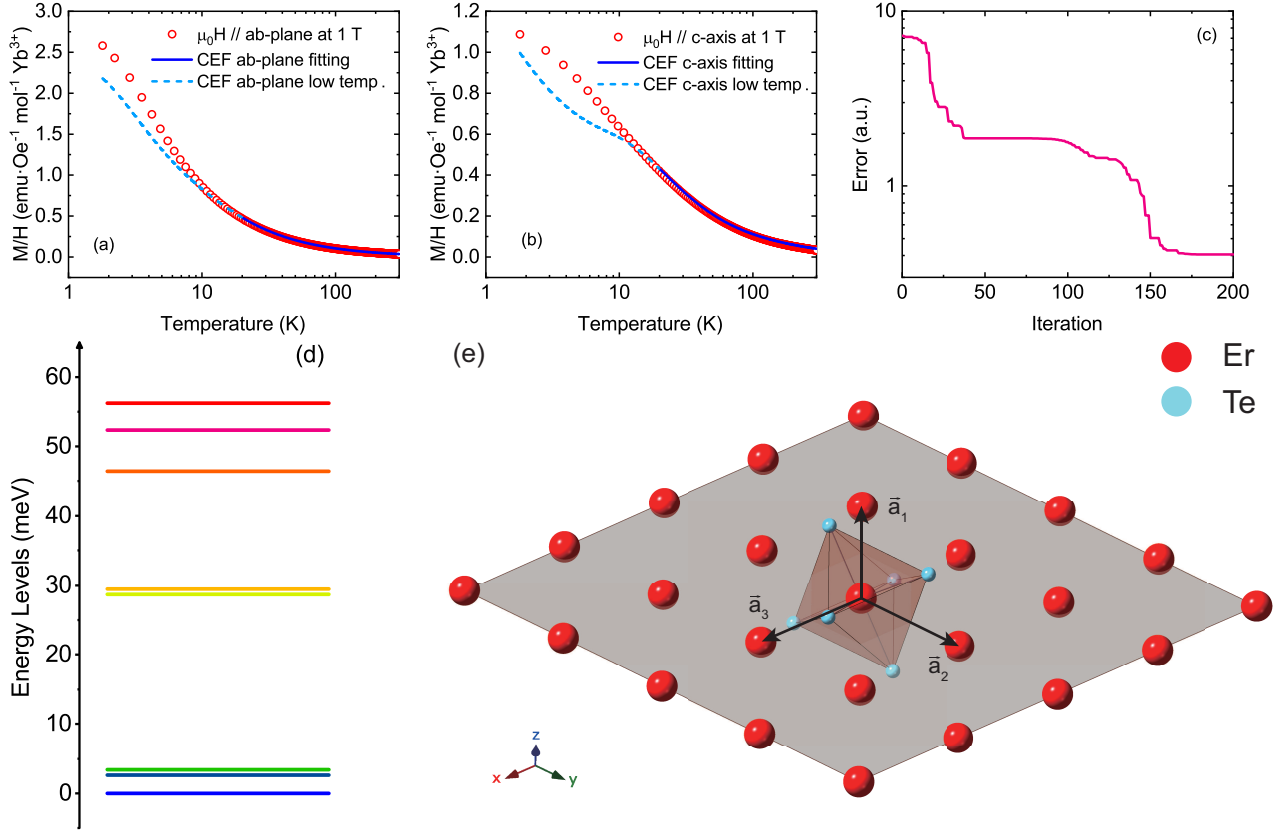}% Here is how to import EPS art
	\caption{\label{fig:epsart}(a) and (b) The susceptibility is obtained when a 1 T magnetic field is applied in the ab-plane and along the c-axis of \ce{KErTe2}, respectively. Red open circles represent the experimental data. The solid line represents fitting results of the CEF susceptibility in the temperature range from 20 to 300 K. The blue dotted line is the CEF calculation result at low temperature. (c) The iteration of the susceptibility fitting. (d) The CEF energy levels of \ce{Er^{3+}} ion. (e) Triangular plane structure composed of \ce{Er^{3+}} in \ce{KErTe2} and the octahedral structure fromed by \ce{Er^{3+}} and \ce{Te^{2-}}.}
\end{figure*}
\emph{Introduction}---Rare-earth chalcogenides \ce{AReCh2} (A = alkali or monovalent metal, Re = rare earth, Ch=O, S, Se, and Te) have recently attracted attention as quantum spin liquid (QSL) candidates\cite{Anderson1973,ANDERSON1987,liu2018rare,bordelon2019field,ding2019gapless,ranjith2019field,baenitz2018naybs,sarkar2019quantum,Zhang2020}. 
Compared with the previous QSL candidates (e.g., \ce{ZnCu3(OH)6Cl2}\cite{PhysRevLett.98.107204}, \ce{EtMe3Sb[Pd(dmit)2]2}\cite{itou2008quantum}, and \ce{YbMgGaO4}\cite{li2015gapless,li2015rare,PhysRevLett.117.097201,PhysRevLett.117.267202,Shen2016,Paddison2016}), most rare-earth chalcogenides do not exhibit only a perfect triangular structure but can also be doped or subjected to element substitution, providing excellent platform for studying strongly correlated electronic systems and complex magnetic interactions\cite{PhysRevMaterials.3.114413,PhysRevB.101.144432}. 
This is closely related to the unique electronic properties of rare-earth ions.
Firstly, strong spin-orbit coupling (SOC) of rare-earth ions induces strong magnetic anisotropy. 
Secondly, Karmers rare-earth ions with an odd number of electrons (e.g., \ce{Er^{3+}} and \ce{Yb^{3+}}) are protected by time-reversal symmetry in the ground state of crystalline electric field (CEF), which is, at least, doubly degenerated. 
Thirdly, unlike the 3d transition metal ions, the 4f electrons of rare-earth ions are shielded by outer electrons, yielding smaller CEF excitations between the ground state and excited states. 
Little CEF energy gap between the ground and excited states critically affected thermodynamic and spectroscopic properties of these materials, which was confirmed in many experimental studies.
\ce{NaYbSe2} exhibited a characteristic temperature of 25 K\cite{Zhang2020}. The characteristic temperature is related to CEF. Above the characteristic temperature, the CEF excitations significantly influenced thermodynamic and spectroscopic results. Furthermore, \ce{KErSe2} has a very small CEF energy gap between the ground state and the 1st excited state ( $\sim$ 1 meV)\cite{PhysRevB.101.144432}, so that susceptibility and magnetization mainly originate from the CEF excitations even at very low temperature.

We have recently successfully synthesized a high-quality single-crystal \ce{KErTe2} sample\cite{liu2021}. 
As the first telluride of rare-earth chalcogenides to be studied, it is isomorphic to \ce{NaYbCh2} (Ch = O, S, Se, and Te) and has a perfect triangular lattice with an R-3m space group. 
The central \ce{Er^{3+}} ion is octahedrally coordinated with six \ce{Te^{2-}} ligands.
Compared with \ce{KErSe2}\cite{PhysRevB.101.144432}, only one element is replaced in \ce{KErTe2}, so we suppose similar CEF excitations of these two compounds. In addition, \ce{KErTe2} has the smallest band energy of $\sim$ 0.9 eV \cite{liu2021} in this family of materials, providing an ideal platform for researching metallization and superconductivities\cite{zhang2020pressure,jia2020mott}. 
Therefore, it is relevant to study the physical properties of \ce{KErTe2} further.

In this paper, we performed the thermodynamic (susceptibility, magnetization, and heat capacity) and spectroscopic measurements (electron spin resonance) of \ce{KErTe2}. 
Based on the measured susceptibility, we successfully fitted the CEF parameters of \ce{KErTe2}.
We calculated the Lande factors using the fitting parameters to verify the accuracy of the fitting CEF parameters, and the results were consistent with those determined in the electron spin resonance (ESR) experiments.
Compared with the traditional method of the CEF excitations measurement based on the inelastic neutron scattering (INS), our method is more cost effective. 
Meanwhile, we fitted and analyzed the susceptibility at low temperature ($\textless$ 10 K) based on the mean-field (MF) theory. Although we tried to fit the susceptibility data with the Curie-Weiss law\cite{liu2021}, the spin-exchange interactions obtained by this method were not accurate and could not explain the effect of CEF on susceptibility and magnetization. By using the self-consistent equation of the MF theory to fit the susceptibility data, we quantitatively determined the magnitude of the anisotropic spin-exchange parameters of \ce{KErTe2}. 
This provided an excellent starting point for further and more advanced calculation methods, such as exact diagonalization (ED) and density matrix renormalization group (DMRG), to study the ground state of \ce{KErTe2}. 
Finally, we measured and analyzed the heat capacity of \ce{KErTe2}. 
%In particular, the observed behavior of the two broad peaks with the different magnetic fields is consistent with the changes we calculated using the CEF theory. Therefore, we conclude that the two broad heat capacity peaks observed in our measurement range are caused by the CEF ground state and excited states.
\begin{figure*}[t]
	\includegraphics[scale=1]{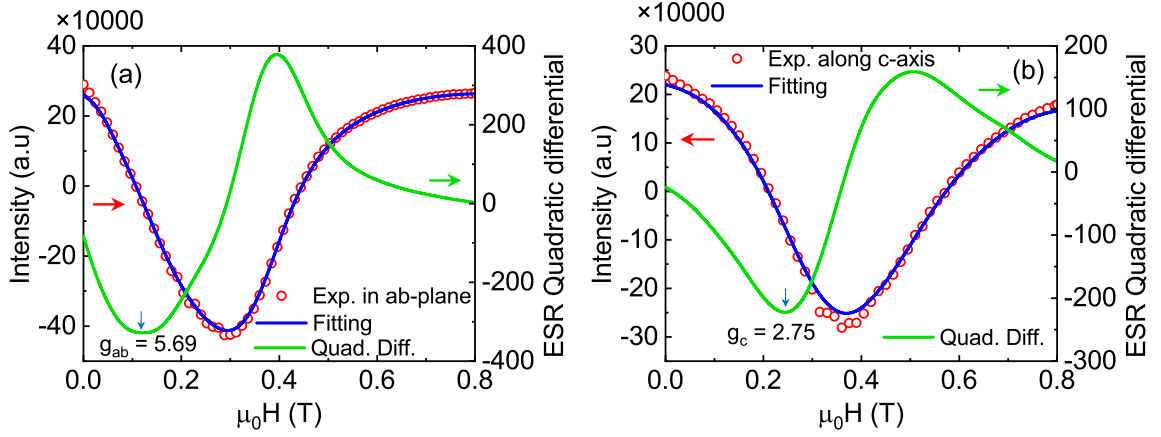}% Here is how to import EPS art
	\caption{\label{fig:epsart}(a) and (b) The ESR spectra of \ce{KErTe2} at 2 K in the ab-plane and along the c-axis. Red open circles represent the experimental results. Solid blue lines are the fitting curves, and solid green lines are the quadratic differential of the fitting curves.}
\end{figure*}

\emph{Samples and Experimental Techniques}---High-quality single crystals of \ce{KErTe2} ($\sim$ 3 mm) were synthesized by using Te-flux method\cite{liu2021}. These single crystals were used for magnetic susceptibility, magnetization, ESR, and heat capacity experiments. Besides \ce{KErTe2}, nonmagnetic isostructural (polycrystalline or single crystal) \ce{KLuTe2} samples\cite{liu2021} were also synthesized in the same way and used for the heat capacity experiments.

About 5 mg  of \ce{KErTe2} single crystals were prepared for the susceptibility and magnetization measurements. 
We performed different measurements along the c-axis and in the ab-plane of \ce{KErTe2} between 1.8 - 300 K at magnetic field of 0 - 14 T.
The heat capacity measurements at 1.8 - 300 K under different magnetic fields were conducted using a Quantum Design Physical Property Measurement System (PPMS). The zero-field heat capacity data of \ce{KErTe2} and \ce{KLuTe2} single crystal samples (about 5 mg) were also measured using the PPMS at temperatures between 1.8 - 100 K. And about 5 mg of \ce{KErTe2} single crystal samples were used for the heat capacity measurements under different magnetic fields parallels to the c-axis of the samples.

The ESR measurements were performed using a Bruker EMX plus 10/12 continuous-wave spectrometer at X-band frequencies (f $\sim$ 9.39 GHz) in the ab-plane and along the c-axis of \ce{KErTe2} at 2 K.

\emph{Electronic configuration, \ce{Er^{3+}} Ion Environment, and Anisotropic Spin Hamiltonian}---The $4f$ orbital of \ce{Er^{3+}} has 11 electrons, so the total spin quantum number $S$ is 3/2. Considering that the orbital quantum number $L$ is 6 and that $4f$ electrons exhibit strong SOC\cite{PhysRevB.94.035107}, there is a spectral term $^{4}I_{15/2}$ with 16-fold degeneracy and a spectral term $^{4}I_{13/2}$ with 14-fold degeneracy. The energy difference between these two spectral terms is about 0.8 eV ($\sim$ 9000 K). Therefore, the spectral term $^{4}I_{15/2}$ represents the ground state of SOC. Moreover, the energy scale of the CEF excitations in rare-earth ions is significantly lower than 0.8 eV, so the influence of the SOC energy level transition on the CEF excitations and magnetism could be neglected.

In \ce{KErTe2}, with the R-3m space group symmetry, the \ce{Er^{3+}} ion and surrounding \ce{Te^{2-}} anions form an octahedral structure with a $D_{3d}$ point group. Therefore, the Hamiltonian of the CEF can be written as follows\cite{li2017crystalline,Zhang2020}:
\begin{equation}
	\hat{H}_{CEF} = \sum_{i} B^{0}_{2}O^{0}_{2} + B^{0}_{4}O^{0}_{4} + B^{3}_{4}O^{3}_{4} + B^{0}_{6}O^{0}_{6} + B^{3}_{6}O^{4}_{6} + B^{6}_{6}O^{6}_{6}
\end{equation}
Among them, $B_{m}^{n}$ is the CEF parameter and $O_{m}^{n}$ is the Steven operator. Since \ce{Er^{3+}} is a Kramers ion composed of an odd number of electrons, each CEF state is doubly degenerated and protected by the time-reversal symmetry. When the measurement temperature is at the same energy scale as the CEF excitation energy level, the CEF excitations significant impact the thermodynamic measurement results. At a lower measurement temperature ($\textless$ 10 K), an electron in the CEF ground state has an effective spin-1/2, and the spin-exchange interaction cannot be ignored. Since \ce{KErTe2} has the same space group as \ce{YbMgGaO4}\cite{li2015rare} and \ce{NaYbSe2}\cite{liu2018rare,Zhang2020}, its anisotropic spin Hamiltonian also has the following form\cite{li2015rare}:
\begin{equation}
\begin{split}
	& \hat{H}_{spin-spin} + \hat{H}_{zeeman} = \\
	& \quad \sum_{\left \langle ij \right \rangle} [J_{zz}S_{i}^{z}S_{j}^{z} + J_{\pm}(S_{i}^{+}S_{j}^{-}+S_{i}^{-}S_{j}^{+}) \\
	& \quad + J_{\pm \pm}(\gamma_{ij}S_{i}^{+}S_{j}^{+}+\gamma_{ij}^{*}S_{i}^{-}S_{j}^{-}) \\
	& \quad -\frac{iJ_{z\pm}}{2}(\gamma_{ij}S_{i}^{+}S_{j}^{z} - \gamma_{ij}^{*}S_{i}^{-}S_{j}^{z} + \left \langle i\longleftrightarrow j \right \rangle)] \\
	& \quad -\mu_{0}\mu_{B}\sum_{i} [g_{ab}(h_{x}S_{i}^{x}+h_{y}S_{i}^{y}) + g_{c}h_{c}S_{i}^{z}]
\end{split}
\end{equation}
where $J_{zz}$, $J_{\pm}$, $J_{\pm\pm}$, and $J_{z\pm}$ are anisotropic spin-exchange parameters, the phase factor $\gamma_{ij}$ = 1, $e^{i2\pi/3}$, $e^{-i2\pi/3}$ for the nearest neighbor (NN) interaction along the $\vec{a}_{1}$, $\vec{a}_{2}$, and $\vec{a}_{3}$ direction, respectively (see Fig. 1(e)). $g_{ab}$ and $g_{c}$ represent the Lande factors from in the ab-plane and along the c-axis, respectively.

Considering that the elemental composition of \ce{KErTe2} and \ce{KErSe2}\cite{PhysRevMaterials.3.114413,PhysRevB.101.144432} are very similar and keeping in mind that the CEF excitation from the gorund state to the 1st excited state is only about 1 meV according to the INS results of \ce{KErSe2}\cite{PhysRevB.101.144432}, the CEF excitations cannot be ignored even at low temperatures. Therefore, we followed the effective Hamiltonian analysis method of \ce{NaYbSe2} at a finite temperature\cite{Zhang2020} when performing the analysis of thermodynamic data of \ce{KErTe2}. The effective Hamiltonian is composed as follows:
\begin{equation}
	\hat{H}_{eff} = \hat{H}_{CEF} + \hat{H}_{spin-spin} + \hat{H}_{zeeman}
\end{equation}
Using the effective Hamiltonian, we fitted the CEF parameters and anisotropic exchange parameters and performed quantitative calculations of the thermodynamic data.
\begin{figure*}[t]
	\includegraphics[scale=1.1]{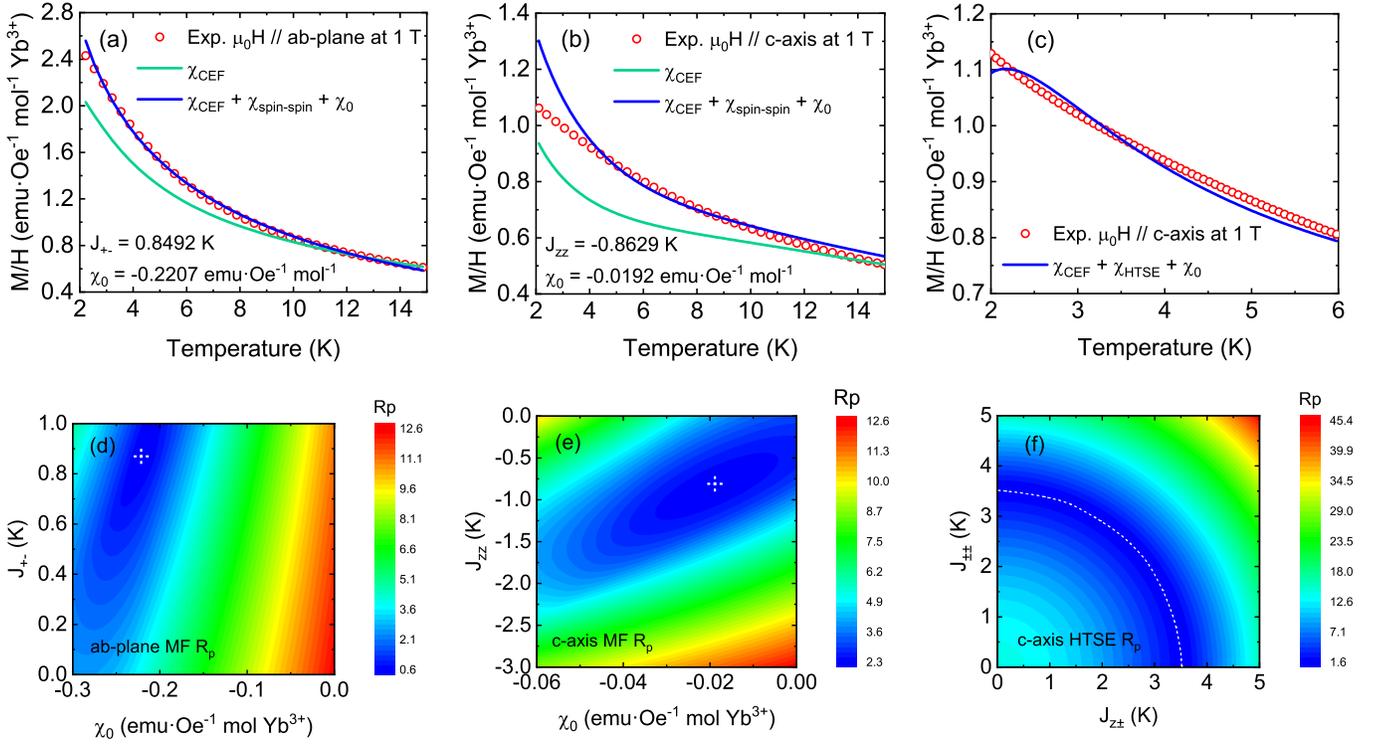}% Here is how to import EPS art
	\caption{\label{fig:epsart}Low-temperature susceptibility MF fitting and spin-spin exchange parameters. (a) and (b) Red open circles, solid blue line, and solid green line are the experimental data, MF fitting, and CEF excitations contribution to susceptibility, respectively. (c) The solid blue line represents the fitting results of HTSE. (d)  The deviation $R_{p}$ of the experimential susceptibility from the fitting results in the ab-plane. (e) and (f) The deviation $R_{p}$ of the experimential susceptibility from the fitting results along the c-axis. White crosses or the white dotted line indicate optimal values.}
\end{figure*}

\emph{Fitting the CEF parameters of \ce{KErTe2}}---Proper understanding of the CEF excitations of rare-earth magnetic materials makes the basis for further research on low-energy quantum magnetism, especially for QSL candidate materials. INS is generally considered the most experimental method to accurately determine the CEF excitations and the related CEF parameters of rare-earth materials. However, INS is quite expensive, and it is not realistic to perform INS measurements on each rare-earth material. Therefore, finding a cost-effective solution for determining the CEF excitations is worth further studying. As far as magnetic susceptibility is concerned, it theoretically contains enough information to analyze the CEF excitations if the measured temperature can cover the CEF excitation energy levels. We can accurately calculate the CEF susceptibility from the CEF Hamiltonian and the thermodynamic calculation formula if we know the CEF parameters. Similarly, we can also fit the CEF parameters by optimizating the algorithm according to the susceptibility\cite{PhysRevB.103.184419}.
The candidate QSL material in this study, \ce{KErTe2}, meets the above requirements. 
Firstly, \ce{KErTe2} and \ce{KErSe2} are very similar. The INS data of \ce{KErSe2}\cite{PhysRevB.101.144432} show that it exhibits small CEF excitations. 
Secondly, our measurement temperature (1.8 $\sim$ 300 K) can partially cover the CEF excitations of \ce{KErTe2}. 
Thirdly, based on the optimization algorithm, we can perform global optimization within the specified interval to find the optimal solution of the fitting parameters.

We choose the susceptibility in the ab-plane and along the c-axis under a magnetic field of 1 T to fit the CEF parameters of \ce{KErTe2}, and the fitting temperature range is 20 - 300 K. We select the CEF parameters of \ce{KErSe2}\cite{PhysRevB.101.144432} as initial parameters, and the fitting interval is [-0.1, 0.1]. The expression used to evaluate the fitting error is given as follows:
\begin{equation}
	R_{p} = \frac{|\chi_{calc} - \chi_{exp}|}{\chi_{exp}}
\end{equation}
The fitting is terminated when the fitting error converges. As shown in Fig. 1(c), when the number of iterations for the parameter fitting is greater than 150, the error function converges to a minimum. The fitted CEF parameters are shown in Table I.
\begin{table*}
	\caption{The fitting results of CEF parameters of \ce{KErTe2}. The units are meV}
	\begin{ruledtabular}
		\begin{tabular}{cccccc}
			$B_{2}^{0}$ & $B_{4}^{0}$ & $B_{4}^{3}$ & $B_{6}^{0}$ & $B_{6}^{3}$ & $B_{6}^{6}$ \\
			\hline
			-0.0764 & 4.6437$\times$10$^{-4}$ & 0.0337 & -4.9049$\times$10$^{-7}$ & -0.9807$\times$10$^{-6}$ & 2.4718$\times$10$^{-6}$
		\end{tabular}
	\end{ruledtabular}
\end{table*}

To verify the accuracy of the fitted CEF parameters, we calculated the Lande factors $g_{ab}$ and $g_{c}$ of \ce{KErTe2}. The calculated results are almost identical to those determined from the ESR experiment (see Fig. 2). The comparison between the calculation and the experimental results is shown in Table II. The calculated results are close to the experimental data, which further verifies the reliability of the fitted CEF parameters. Since the g factor is an important indicator of material anisotropy, it is necessary to make a simple comparison with the g factor of other QSL candidate materials. Based on the existing experimental data, we found that the magnetic anisotropy of \ce{AReCh2} is diverse. \ce{NaYbS2}\cite{sichelschmidt2019electron} and \ce{NaYbSe2}\cite{Sichelschmidt2020,Zhang2020} compounds, where \ce{Yb^{3+}} is a magnetic ion, have the Lande factor ratios of $g_{ab}$: $g_{c}$ = 3: 1, showing stronger anisotropy than \ce{YbMgGaO4} ($g_{ab}$: $g_{c}$ = 1.2: 1)\cite{li2015rare}. Among them, \ce{Na^{+}} ions play a key role in the anisotropy\cite{zangeneh2019single}. Besides, the Lande factor ratios of $g_{ab}$: $g_{c}$ are 1.4: 1 and 2: 1 in \ce{KErSe2}\cite{PhysRevB.101.144432} and \ce{KErTe2}, respectively, suggesting that not only alkali metal ions but also chalcogenide anions can affect the magnetic anisotropy.
\begin{table}
	\caption{Comparison of the Experimental and Calculated Results of the Lande Factors}
	\begin{ruledtabular}
		\begin{tabular}{ccc}
			               & $g_{ab}$ & $g_{c}$  \\
			\hline
			ESR Experiment & 5.69 & 2.75         \\
			\hline
			CEF Calculation & 5.89(7) & 2.73(4) 
		\end{tabular}
	\end{ruledtabular}
\end{table}

Furthermore, we calculated the susceptibility in the ab-plane and along the c-axis of \ce{KErTe2} under a 1 T magnetic field in the temperature range of 1.8 - 20 K using these parameters. The calculated results are shown in Fig. 1(a) and (b), respectively. By comparing experimental results, the susceptibility data are consistent with the calculated results above 10 K. At low temperature ($\textless$ 10 K), the susceptibility calculated using only the CEF theory exhibits a significant deviation from the experimental data. In this temperature range, the spin-exchange interaction is no longer little so that the susceptibility contributed by the spin-exchange interaction cannot be ignored.

We calculated every CEF energy level of \ce{KErTe2}, Fig. 1(d), based on the fitting CEF parameters. The 1st (2.65 meV) and the 2nd (3.43 meV) excited energy levels are smaller, and the temperature of susceptibility measurement does not exceed 300 K, so these CEF excited states significantly affect the susceptibility.

\begin{figure*}[t]
	\includegraphics[scale=1]{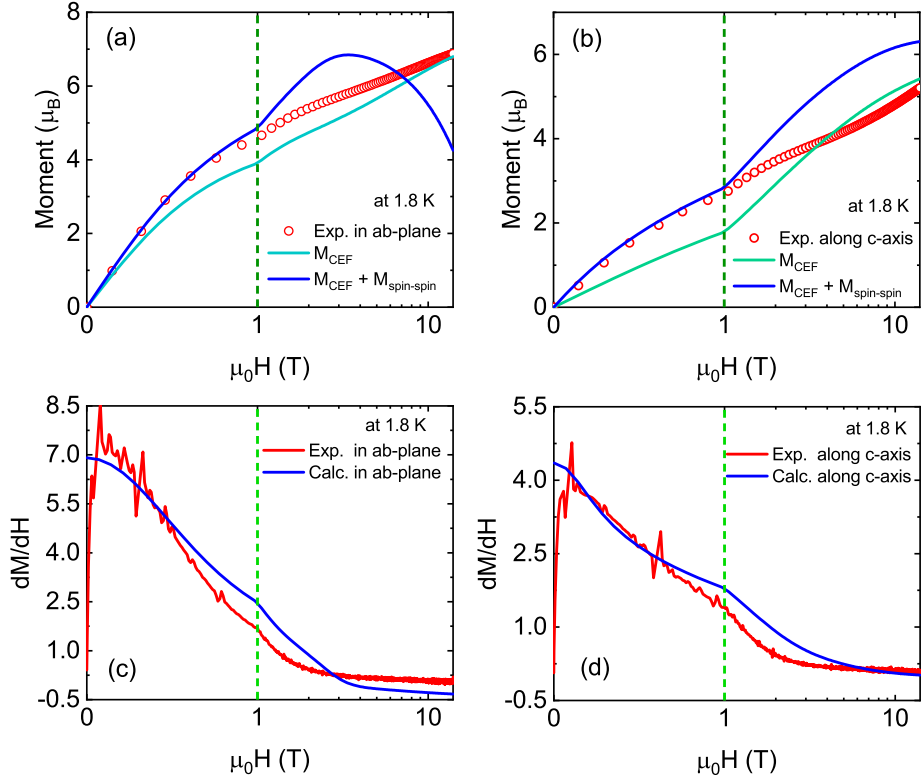}% Here is how to import EPS art
	\caption{\label{fig:epsart} (a) and (b) The magnetizations curves in the ab-plane and along the c-axis at 1.8 K, respectively. Red open circles represent the experimential results, and the solid deep blue lines and light blue lines are the calculated results based on the MF theory and CEF theory, respectively. Their corresponding derivatives are shown in (c) and (d).}
\end{figure*}
\emph{Susceptibility, Magnetization and Mean Field (MF) Fitting}---For the susceptibility in a low-temperature range ($\textless$ 10 K), it is not enough to consider only the contribution of the CEF to magnetism. As shown in Fig. 1(a) and Fig. 1(b), the susceptibility of the CEF deviates significantly from the experimental data below 10 K. To correct the calculation results of the low temperature susceptibility, we have considered the correction under the MF approximation of the spin-spin Hamiltonian. As shown in Fig. 3 (a) and (b), we select 1.8 - 15 K as the fitting temperature range, and add the MF contribution of spin-exchange interactions to susceptibility. Consequently, the calculation results are closer to the experimental measurement results.  
Besides, we also derive the two spin-exchange parameters $J_{\pm}$ and $J_{zz}$, which are 0.85 and -0.86 K, respectively. $\chi_{0}$ in the ab-plane and along c-axis are -0.2207 and -0.0192 $emu \cdot Oe^{-1} \cdot mol^{-1}$, respectively. 
It should be noted that the fitting results deviate from the experimental data less than 5 K in the c-axis direction, plausibly because the $J_{\pm\pm}$ and $J_{z\pm}$ terms in the spin Hamiltonian are ignored when we use the MF approximation. The high-temperature series expansion (HTSE) can give the range of the other two anisotropic exchange parameters $J_{\pm\pm}$ and $J_{z\pm}$. The susceptibility formula of the HTSE for the c-axis is\cite{li2015rare,Zhang2020}
\begin{equation}
	\chi_{c} = \frac{\mu_{0}g_{c}^{2}\mu_{B}^{2}}{4k_{B}T}\left( 1-\frac{3J_{zz}}{2k_{B}T} - \frac{3J_{\pm}^{2} + J_{\pm\pm}^{2} + J_{z\pm}^{2}}{2k_{B}^{2}T} + \frac{15J_{zz}^{2}}{8k_{B}^{2}T^{2}} \right)
\end{equation}
We choose the c-axis susceptibility data below 6 K to fit $J_{\pm\pm}$ and $J_{z\pm}$, and the finally determined the range of $J_{\pm\pm}^{2} + J_{z\pm}^{2} = 12.48$ K$^{2}$ (see Fig. 3(c)), indicating that the anisotropic interaction may be large in \ce{KErTe2}, significantly affecting the susceptibility in the c-axis direction.

The magnetization measured by the PPMS at 1.8 K is illustrated in Fig. 4(a) and 4(b). The magnetization contributed by the CEF and spin-exchange interactions with the MF approximation is also calculated. Independently of the ab-plane or the c-axis magnetizations, the calculated results are significantly different from the experiment data when the CEF contribution is considered. When we add spin-exchange interactions, the calculated results are consistent with the experimental data below 1 T. However, even when the spin-exchange interactions are considered, the experimental data and the calculated results are significantly different from those above 1 T. There may be three various reasons for this behavior. Firstly, similar to the problem related to the susceptibility calculation, the other two exchange parameters $J_{\pm\pm}$ and $J_{z\pm}$ are ignored under the MF approximation, causing a significant deviation of the calculation result from the experimental data under high field magnetic fields. Secondly, by fitting the magnetic susceptibility, we know that the first and the second excitation energy levels of the CEF of \ce{KErTe2} are only a few meV. When the applied magnetic field is high, it is easy to split the doubly degenerated CEF energy level (similar to the Zeeman splitting), affecting the magnetism of \ce{KErTe2}. This phenomenon is observed in the measurement of the \ce{KErTe2} heat capacity with the magnetic field, which will be discussed later in detail. 
Thirdly, \ce{KErTe2} may undergo a magnetic field-induced phase transformation at a higher magnetic fields. A small but obvious kink can be observed near 1 T, especially in the dM/dT-H curve (Fig. 4(c) and Fig. 4(d)). Field-induced phase transitions are alos very common in other rare-earth chalcogenide QSL candidate compounds, e.g., in the dM/dH curves or AC susceptibility of \ce{NaYbO2}\cite{bordelon2019field}, \ce{NaYbS2}\cite{ma2020spin}, and \ce{NaYbSe2}\cite{ranjith2019anisotropic,Zhang2020}.

\begin{figure*}[t]
	\includegraphics[scale=0.9]{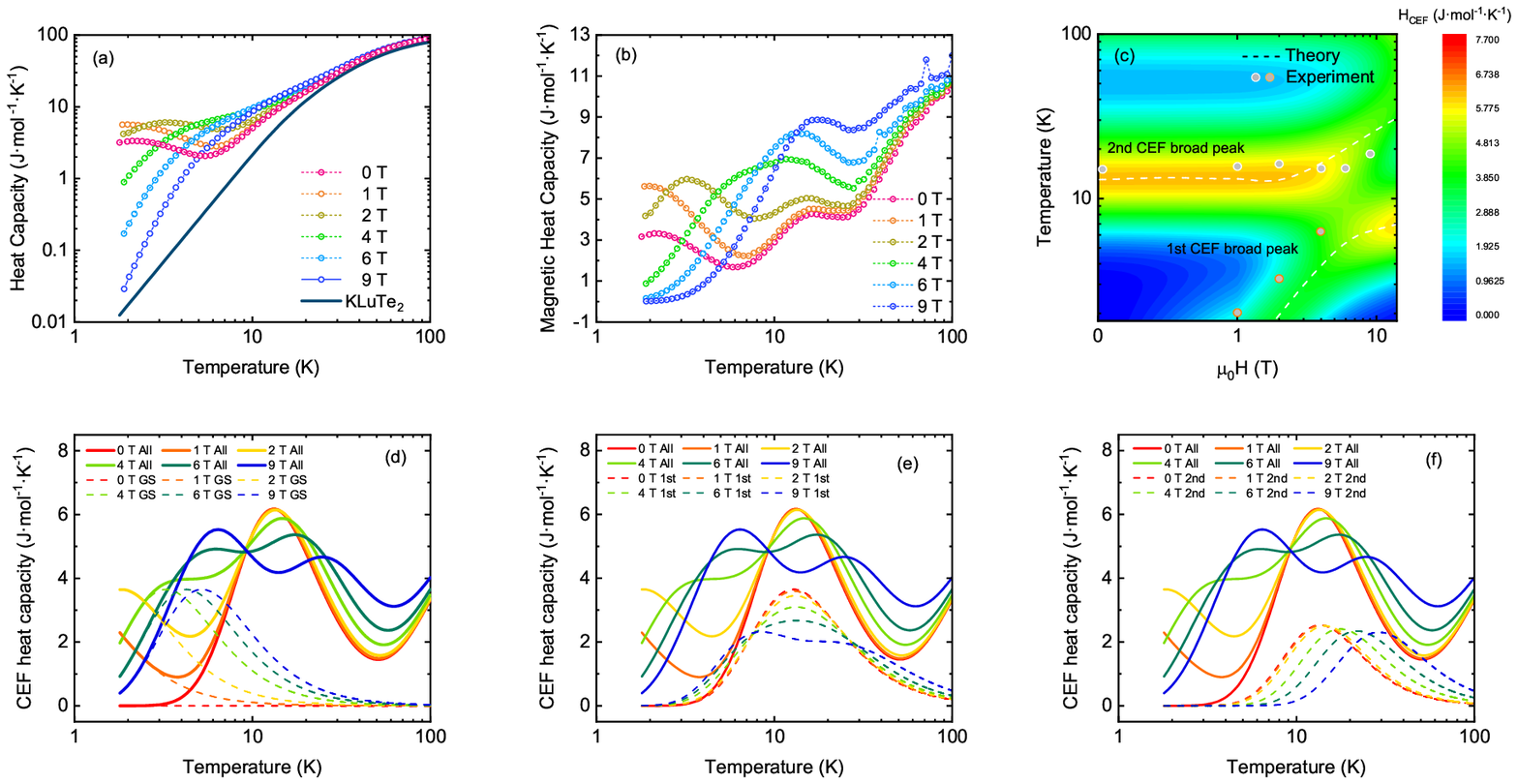}% Here is how to import EPS art
	\caption{\label{fig:epsart} Experimental results of the \ce{KErTe2} heat capacity at different magnetic fields along the c-axis and the zero field heat capacity of \ce{KLuTe2}. (b) The magnetic heat capacity of \ce{KErTe2}. \ce{KLuTe2} is used to simulate the lattice heat capacity of \ce{KErTe2}. (c) The contour of the \ce{KErTe2} CEF heat capacity. White dotted lines mark the calculated central peak positions of the CEF heat capacity, and gray balls represent the observed central peak position of the \ce{KErTe2} magnetic heat capacity. (d), (e), and (f) The heat capacity calculated results under different magnetic fields based on the CEF theory and thermodynamic calculation method. The solid line is the contribution of all CEF states. Dotted lines in (d), (e), and (f) represent the contributions of the ground state, 1st excited state, and 2nd excited state to the CEF heat capacity, respectively.}
\end{figure*}
\emph{Heat Capacity}---A relatively small energy gap between the ground state and the 1st or 2nd excitation states of CEF in \ce{KErTe2} will also have a significant impact on the heat capacity measurement. This was also reflected in the zero-field heat capacity measurement of \ce{NaYbSe2}\cite{ranjith2019anisotropic,Zhang2020,dai2020spinon}. To confirm this and further study the influence of the CEF excitations on the heat capacity at different magnetic fields, we measured the heat capacity of \ce{KErTe2} at zero-field and different magnetic fields. As shown in Fig. 5(a), the heat capacity of \ce{KErTe2} and the nonmagnetic isostructural samples of \ce{KLuTe2} at different magnetic fields are measured at temperatures from 1.8 to 100 K. \ce{KLuTe2} is used to simulate the contribution of the lattice heat capacity. 
The magnetic heat capacity of \ce{KErTe2}, Fig. 5(b), is obtained by subtracting the lattice heat capacity simulated by \ce{KLuTe2}. We can observe two broad peaks at $\sim$ 2 and $\sim$ 10.8 K in the zero-field magnetic heat capacity. With the increase of the magnetic field, both peaks shift to the high-temperature direction. The low-temperature peak overlaps with the high-temperature peak at 4 T. Based on the previous analysis\cite{liu2021}, both peaks may be related to the CEF excitations. The 1st and the 2nd excitation energy levels of the CEF of \ce{KErTe2} are calculated to be very close to the energy scales that correspond to the positions of these two peaks. Thus, we have calculated the heat capacity contribution by the CEF excitations of \ce{KErTe2} based on the fitted CEF parameters. At the zero-field conditions (see Fig. 5(d)), the CEF contribution to the heat capacity is almost zero below 4 K. However, the experimental magnetic heat capacity of \ce{KErTe2} is not zero at the zero-field conditions (see Fig. 5(b)). This shows that the magnetic heat capacity mainly originates from spin-exchange correlation interactions in the low-temperature range ($\textless$ 4 K) and zero-field conditions, and it is close to the spin-exchange parameters ($\sim$ 0.8 K) obtained by fitting the susceptibility.

When the magnetic field is increased to 1 T, the experimental and calculation results show that the magnetic heat capacity significantly increases at 1.8 K. Both increments are about 3.5 $J \cdot mol^{-1} \cdot K^{-1}$, which may be related to the splitting of the CEF ground state under an external magnetic field. To confirm this, we have calculate the heat capacity contributed by the CEF ground state under different magnetic fields. The calculated data are shown in Fig. 5(d) and marked by dotted lines. At zero-field conditions, the CEF ground state does not contribute to the heat capacity. At 1 T, the CEF ground state is split, and the increment in heat capacity is 3.5 $J \cdot mol^{-1} \cdot K^{-1}$ at 1.8 K, which is completely consistent with our assumption. 
Besides, the broad peak at about 10.8 K mainly originates from the 1st and the 2nd excited states of the CEF. We have calculated the heat capacity contributed by the 1st and the 2nd excited states of CEF, which are marked by dotted lines in Fig. 5(e) and (f). Unlike the CEF ground state, the 1st and the 2nd excited states of CEF contribute to the heat capacity even at zero-field.
As the magnetic field increases, the heat capacity peak corresponding to the contribution of the 1st excited state becomes broader, and the 2nd excited state shifts to the high-temperature direction. 
Combining the CEF energy levels calculated before, this is easy to understand. First, the  magnetic field splits the doubly degenerated CEF energy level, so the magnetic heat capacity peak is broadened. Second, the 3rd CEF energy level is much higher than the 1st and the 2nd  CEF levels. Within our measurement temperature range, particles can handly exceed the 2nd CEF energy level. The magnetic moment contributed by the 2nd CEF energy level will be polarized in the same direction by the magnetic field (similar to the arrangement of the ferromagnetic magnetic moment). Therefore, as the magnetic field increases, the magnetic heat capacity peak shifts in the high temperature direction.
Comparing the changes of the two peaks with the magnetic field in detail, we plotted the contour of the \ce{KErTe2} CEF heat capacity, Fig. 5(c). Two white dotted lines mark the changing trend of the center peak position of the two broad peaks contributed by the CEF with the magnetic field. Gray balls represent the observed central peak position of the \ce{KErTe2} magnetic heat capacity. These balls are almost distributed along the two dotted lines. It is worth noting that we have only calculated the heat capacity contributed by the CEF, although the heat capacity in the experimental includes the contribution besides the CEF. Especially at low temperatures, the interaction between spins gradually begins to play a leading role, so our calculated results cannot be completely consistent with experimental results. Still, our calculated results are globally consistent with the experimental results.

\emph{Summary}---We used the effective Hamiltonian\cite{Zhang2020} to perform detailed fitting, calculation, and analysis of the thermodynamic data of \ce{KErTe2} at a finite temperature. Firstly, using the high-temperature susceptibility data, we successfully fitted the CEF parameters of \ce{KErTe2}, and the g factor obtained by using these parameters was consistent with that determined from the ESR experiments, confirming accuracy of the used parameters. Secondly, we proved that the CEF excitations in \ce{KErTe2} played a decisive role in the magnetism above 3 K, while the magnetism below 10 K corresponded to the anisotropic spin-exchange  interactions. Based on the CEF theory and MF approximation of the spin-spin Hamiltonian, we fitted and calculated the susceptibility and magnetization data below the characteristic temperature. Meanwhile, we obtained the spin-spin exchange parameters $J_{\pm}$  and $J_{zz}$. Under the HTSE, we provided a range of other anisotropic exchange parameters $J_{\pm\pm}$ and $J_{z\pm}$. Thirdly, we successfully simulated the change in two peaks of the magnetic heat capacity with the magnetic field by using the CEF theory and thermodynamic formulas. Among them, the ground state, and the 1st and 2nd excited states of CEF, and the splitting of these states under a magnetic field represented the main contributions factors to the magnetic heat capacity at low temperatures.

\emph{Acknowledgments}---This work was supported by the National Key Research and Development Program of China (Grant No. 2017YFA0302904), the National Science Foundation of China (Grant Nos. U1932215 and No. 11774419), and the Strategic Priority Research Program of the Chinese Academy of Sciences (Grant No. XDB33010100). Q.M.Z. acknowledges the support from Users with Excellence Program of Hefei Science Center and High Magnetic Field Facility, CAS. All the fittings and calculations are based on the program package MagLab v0.2, which is designed for the fitting, calculation, and analysis of magnetism. The code has been further extended and optimized since the version of v0.1 and the new version is developed based on MathWorks MATLAB software (Academic License for Renmin University of China). The authors would like to express their gratitude to EditSprings (https://www.editsprings.com/) for the expert linguistic services provided.

\end{document}